\documentclass[preprint]{aastex}

\newcommand{\yr}{{\rm\,yr}}

\newcommand{\au}{{\rm\,AU}}
\newcommand{\mps}{{\rm\,m\,s}^{-1}}

\begin{document}

\title{On the 2:1 Orbital Resonance in the HD 82943 Planetary
       System\footnote{
       Based on observations obtained at the W. M. Keck Observatory,
       which is operated jointly by the University of California and
       the California Institute of Technology.}
}
\author{Man Hoi Lee\altaffilmark{2}, R. Paul Butler\altaffilmark{3},
        Debra A. Fischer\altaffilmark{4}, Geoffrey W.
        Marcy\altaffilmark{5}, and Steven S. Vogt\altaffilmark{6}}
\altaffiltext{2}{Department of Physics, University of California,
                 Santa Barbara, CA 93106.}
\altaffiltext{3}{Department of Terrestrial Magnetism, Carnegie
                 Institution of Washington, 5241 Broad Branch Road NW,
                 Washington, DC 20015-1305.}
\altaffiltext{4}{Department of Physics and Astronomy, San Francisco
                 State University, San Francisco, CA 94132.}
\altaffiltext{5}{Department of Astronomy, University of California,
                 Berkeley, CA 94720.}
\altaffiltext{6}{UCO/Lick Observatory, University of California, Santa
                 Cruz, CA 95064.}

\begin{abstract}
We present an analysis of the HD 82943 planetary system based on a
radial velocity data set that combines new measurements obtained with
the Keck telescope and the CORALIE measurements published in graphical
form.
We examine simultaneously the goodness of fit and the dynamical
properties of the best-fit double-Keplerian model as a function of the
poorly constrained eccentricity and argument of periapse of the
outer planet's orbit.
The fit with the minimum $\chi_{\nu}^2$ is dynamically unstable if the
orbits are assumed to be coplanar.
However, the minimum is relatively shallow, and there is a wide range
of fits outside the minimum with reasonable $\chi_{\nu}^2$.
For an assumed coplanar inclination $i = 30^{\circ}$ ($\sin i = 0.5$),
only good fits with both of the lowest order, eccentricity-type
mean-motion resonance variables at the 2:1 commensurability,
$\theta_1$ and $\theta_2$, librating about $0^{\circ}$ are stable.
For $\sin i = 1$, there are also some good fits with only $\theta_1$
(involving the inner planet's periapse longitude) librating that are
stable for at least $10^8$ years.
The libration semiamplitudes are about $6^{\circ}$ for $\theta_1$ and
$10^{\circ}$ for $\theta_2$ for the stable good fit with the smallest
libration amplitudes of both $\theta_1$ and $\theta_2$.
We do not find any good fits that are non-resonant and stable.
Thus the two planets in the HD 82943 system are almost certainly in
2:1 mean-motion resonance, with at least $\theta_1$ librating, and the
observations may even be consistent with small-amplitude librations of
both $\theta_1$ and $\theta_2$.
\end{abstract}

\section{INTRODUCTION}

The first pair of extrasolar planets suspected to be in mean-motion
resonance was discovered around the star GJ 876, with the orbital
periods nearly in the ratio 2:1 \citep{mar01}.
A dynamical fit to the radial velocity data of GJ 876 that accounts
for the mutual gravitational interaction of the planets is essential
because of the short orbital periods ($\approx 30$ and 60 days) and
large planetary masses [minimum combined planetary mass relative to
the stellar mass $(m_1 + m_2)/m_0 \approx 0.0074$].
It is now well established that this pair of planets is deep in 2:1
orbital resonance, with both of the lowest order, eccentricity-type
mean-motion resonance variables,
\begin{equation}
\theta_1 = \lambda_1 - 2\lambda_2 + \varpi_1
\label{theta1}
\end{equation}
and
\begin{equation}
\theta_2 = \lambda_1 - 2\lambda_2 + \varpi_2 ,
\label{theta2}
\end{equation}
librating about $0^\circ$ with small amplitudes
\citep{lau01,riv01,lau05}.
Here $\lambda_{1,2}$ are the mean longitudes of the inner and outer
planets, respectively, and $\varpi_{1,2}$ are the longitudes of
periapse.
The simultaneous librations of $\theta_1$ and $\theta_2$ about
$0^\circ$ mean that the secular apsidal resonance variable,
\begin{equation}
\theta_{\rm SAR} = \varpi_1 - \varpi_2 = \theta_1 - \theta_2 ,
\label{thetaSAR}
\end{equation}
also librates about $0^\circ$ and that the periapses are on average
aligned.
The basic dynamical properties of this resonant pair of planets are
not affected by the recent discovery of a third, low-mass planet on a
1.9-day orbit in the GJ 876 system \citep{riv05}.

It is important to confirm other suspected resonant planetary systems,
as the GJ 876 system has shown that resonant planetary systems are
interesting in terms of both their dynamics and their constraints on
processes during planet formation.
The geometry of the 2:1 resonances in the GJ 876 system is different
from that of the 2:1 resonances between the Jovian satellites Io and
Europa, where $\theta_1$ librates about $0^\circ$ but $\theta_2$ and
$\theta_{\rm SAR}$ librate about $180^\circ$.
For small orbital eccentricities, the Io-Europa configuration is the
only stable 2:1 resonance configuration with both $\theta_1$ and
$\theta_2$ librating.
For moderate to large eccentricities, the Io-Europa configuration is
not stable, but there is a wide variety of other stable 2:1 resonance
configurations, including the GJ 876 configuration
\citep{lee02,lee03a,bfm03,fbm03,lee04}.
The resonances in the GJ 876 system were most likely established by
converging differential migration of the planets leading to capture
into resonances, with the migration due to interaction with the
protoplanetary disk.
While it is easy to establish the observed resonance geometry of the
GJ 876 system by convergent migration, the observational upper limits
on the eccentricities ($e_1 \la 0.31$ and $e_2 \la 0.05$) require
either significant eccentricity damping from planet-disk interaction
or resonance capture occurring just before disk dispersal, because
continued migration after resonance capture could lead to rapid growth
of the eccentricities \citep{lee02,kle05}.
Hydrodynamic simulations of the assembly of the GJ 876 resonances
performed to date do not show significant eccentricity damping from
planet-disk interaction and produce eccentricities that exceed the
observational upper limits, unless the disk is dispersed shortly after
resonance capture \citep{pap03,kle04,kle05}.

HD 82943 was the second star discovered to host a pair of planets with
orbital periods nearly in the ratio 2:1.
The discovery of the first planet was announced in ESO Press Release
13/00\footnote{
See
\anchor{http://www.eso.org/outreach/press-rel/pr-2000/pr-13-00.html}
{http://www.eso.org/outreach/press-rel/pr-2000/pr-13-00.html}
}
and the discovery of the second, inner planet was announced in ESO
Press Release 07/01\footnote{
See
\anchor{http://www.eso.org/outreach/press-rel/pr-2001/pr-07-01.html}
{http://www.eso.org/outreach/press-rel/pr-2001/pr-07-01.html}
}.
\citet{may04} have recently published the radial velocity measurements
obtained with the CORALIE spectrograph on the 1.2-m Euler Swiss
telescope at the ESO La Silla Observatory in graphical form only.
Unlike GJ 876, a double-Keplerian fit is likely adequate for HD 82943,
because the orbital periods are much longer ($\approx 220$ and 440
days) and the planetary masses are smaller [minimum $(m_1 + m_2)/m_0
\approx 0.003$], and the mutual gravitational interaction of the
planets is not expected to affect the radial velocity significantly
over the few-year time span of the available observations.
\citet{may04} have found a best-fit double-Keplerian solution, and its
orbital parameters are reproduced in Table \ref{tab:mayor}.

\citet{fmb05} have reported simulations of the \citet{may04} best-fit
solution that are unstable, but they assumed that the orbital
parameters are in astrocentric coordinates and they did not state the
ranges of orbital inclinations and starting epochs examined.
The fact that the orbital parameters obtained from multiple-Keplerian
fits should be interpreted as in Jacobi coordinates was first
pointed out by \citet{lis01} and was derived and demonstrated by
\citet{lee03b}.
We have performed direct numerical orbit integrations of the
\citet{may04} best-fit solution, assuming that the orbital parameters
are in Jacobi coordinates and that the orbits are coplanar with the
same inclination $i$ from the plane of the sky.
When we assumed that the orbital parameters correspond to the
osculating parameters at the epoch JD 2451185.1 of the first CORALIE
measurement, the system becomes unstable after a time ranging from
$\sim 500$ to $4 \times 10^4\yr$ for $\sin i = 1$, $0.9$, $\ldots$,
$0.5$.
Since the long-term evolution can be sensitive to the epoch that the
orbital parameters are assumed to correspond to, we have repeated the
direct integrations by starting at three other epochs equally spaced
between the first (JD 2451185.1) and last (JD 2452777.7) CORALIE
measurements.
All of the integrations become unstable, with the vast majority in
less than $10^6\yr$.
These results and those in \citet[][who also examined mutually
inclined orbits and the effects of the uncertainty in the stellar
mass]{fmb05} show that the best-fit solution found by \citet{may04} is
unstable.

It is, however, essential that one examines not only the best-fit
solution that minimizes the reduced chi-square statistic $\chi_\nu^2$,
but also fits with $\chi_\nu^2$ not significantly above the minimum,
especially if the $\chi_\nu^2$ minimum is shallow and $\chi_\nu^2$
changes slowly with variations in one or more of the parameters.
\citet{fmb05} have analyzed the CORALIE data published in
\citet{may04} by generating a large number of orbital parameter sets
and selecting those that fit the radial velocity data with the RMS of
the residuals (instead of $\chi_\nu^2$) close to the minimum value.
The stable coplanar fits that they have found have arguments of
periapse $\omega_1 \approx 120^\circ$ and $\omega_2 \ga 200^\circ$,
and both $\theta_1$ and $\theta_2$ librate about $0^\circ$ with large
amplitudes.\footnote{
For coplanar orbits, the longitudes of periapse $\varpi_j$ in the
resonance variables $\theta_1$, $\theta_2$, and $\theta_{\rm SAR}$
(eqs. [\ref{theta1}]--[\ref{thetaSAR}]) can be measured from any
reference direction in the orbital plane.
If we choose the ascending node referenced to the plane of the sky as
the reference direction, $\varpi_j$ are the same as the arguments of
periapse $\omega_j$ obtained from the radial velocity fit.
}
On the other hand, \citet{may04} have noted that it is possible to
find an aligned configuration with $\omega_1 \approx \omega_2$ that
fits the data with nearly the same RMS as their best-fit solution,
although it is unclear whether such a fit would be stable and/or in
2:1 orbital resonance.
Thus, it has not been firmly established that the pair of planets
around HD 82943 are in 2:1 orbital resonance, even though it is likely
that a pair of planets of order Jupiter mass so close to the 2:1
mean-motion commensurability would be dynamically unstable unless they
are in 2:1 orbital resonance.

In this paper we present an analysis of the HD 82943 planetary system
based on a radial velocity data set that combines new measurements
obtained with the Keck telescope and the CORALIE measurements
published in graphical form.
The stellar characteristics and the radial velocity measurements are
described in \S 2.
In \S 3 we present the best-fit double-Keplerian models on a grid of
the poorly constrained eccentricity, $e_2$, and argument of periapse,
$\omega_2$, of the outer planet's orbit.
In \S 4 we use dynamical stability to narrow the range of reasonable
fits and examine the dynamical properties of the stable fits.
We show that the two planets in the HD 82943 system are almost
certainly in 2:1 orbital resonance, with at least $\theta_1$
librating, and may even be consistent with small-amplitude librations
of both $\theta_1$ and $\theta_2$.
Our conclusions are summarized and discussed in \S 5.

\section{STELLAR CHARACTERISTICS AND RADIAL VELOCITY MEASUREMENTS}

The stellar characteristics of the G0 star HD 82943 were summarized in
\citet{may04} and \citet{fis05}.
\citet{san00a} and \citet{law03} have determined a stellar mass of
$m_0 = 1.08 M_\odot$ and $1.18^{+0.05}_{-0.01} M_\odot$, respectively.
\citet{fis05} have measured the metallicity of HD 82943 to be
${\rm [Fe/H]} = +0.27$ which, when combined with evolutionary models,
gives a stellar mass of $1.18 M_\odot$.
We follow \citet{may04} and adopt a mass of $m_0 = 1.15 M_\odot$.

Stellar radial velocities exhibit intrinsic velocity ``jitter'' caused
by acoustic p-modes, turbulent convection, star spots, and flows
controlled by surface magnetic fields.
The level of jitter depends on the age and activity level of the
star \citep{saa98,cum99,san00b,wri05}.
From the observed velocity variance of hundreds of stars with no
detected companions in the California and Carnegie Planet Search
Program, \citet{wri05} has developed an empirical model for predicting
the jitter from a star's $B-V$ color, absolute magnitude, and
chromospheric emission level.
Some of this empirically estimated ``jitter'' is no doubt actually
instrumental, stemming from small errors in the data analysis at
levels of 1--$2\mps$.
Based on the model presented in \citet{wri05}, we estimate the jitter
for HD 82943 to be $4.2\mps$, with an uncertainty of $\sim 50\%$.

We began radial velocity measurements for HD 82943 in 2001 with the
HIRES echelle spectrograph on the Keck I telescope.
We have to date obtained 23 velocity measurements spanning $3.8\yr$,
and they are listed in Table \ref{tab:RV}.
The median of the internal velocity uncertainties is $3.0\mps$.

\citet{may04} have published 142 radial velocity measurements obtained
between 1999 and 2003 in graphical form only.
We have extracted their data from Figure 6 of \citet{may04} with an
accuracy of about $0.5\,$d in the times of observation and an accuracy
of about $0.07\mps$ in the velocities and their internal
uncertainties.
The uncertainty of $0.5\,$d in the extracted time of observation can
translate into a velocity uncertainty of as much as $2\mps$ in the
steepest parts of the radial velocity curve (where the radial velocity
can change by as much as $4\mps$ per day), but the square of even this
maximum ($2\mps$) is much smaller than the sum of the squares of the
extracted internal uncertainty ($\ga 4\mps$) and the estimated stellar
jitter ($4.2\mps$).
As another check of the quality of this extracted CORALIE data set, we
have performed a double-Keplerian fit of the extracted data, with the
extracted internal uncertainties as the measurement errors in the
calculation of the reduced chi-square statistic $\chi_\nu^2$.
The resulting best fit has RMS $= 6.99\mps$, which is only slightly
larger than the RMS ($= 6.8\mps$) of the Mayor et al. best fit, and
the orbital parameters agree with those found by Mayor et al. and
reproduced in Table \ref{tab:mayor} to better than $\pm 2$ in the last
digit.
This confirms that the extracted data set is comparable in quality to
the actual data set.

In the analysis below, we use the combined data set containing 165
velocity measurements and spanning $6.1\yr$, and we adopt the
quadrature sum of the internal uncertainty and the estimated stellar
jitter as the measurement error of each data point in the calculation
of $\chi_\nu^2$.

\section{DOUBLE-KEPLERIAN FITS}

As we mention in \S 1, the mutual gravitational interaction of the
planets is not expected to affect the radial velocity of HD 82943
significantly over the $6.1$-year time span (or about 5 outer planet
orbits) of the available observations.
Thus we fit the radial velocity $V_r$ of the star with the reflex
motion due to two orbiting planetary companions on independent
Keplerian orbits:
\begin{equation}
V_r = K_1 \left[\cos(\omega_1 + f_1) + e_1 \cos\omega_1\right] +
      K_2 \left[\cos(\omega_2 + f_2) + e_2 \cos\omega_2\right] + V ,
\label{RV}
\end{equation}
where $V$ is the line-of-sight velocity of the center of mass of the
whole planetary system relative to the solar system barycenter, $K_j$
is the amplitude of the velocity induced by the $j$th planet, and
$e_j$, $\omega_j$ and $f_j$ are, respectively, the eccentricity,
argument of periapse, and true anomaly of the $j$th planet's orbit
(see, e.g., \citealt{lee03b}).
The zero points of the velocities measured by different telescopes are
different, and $V$ is a parameter for each telescope (denoted as $V_K$
and $V_C$ for the Keck and CORALIE data, respectively).
Since the true anomaly $f_j$ depends on $e_j$, the orbital period
$P_j$, and the mean anomaly ${\cal M}_j$ at a given epoch (which we
choose to be the epoch of the first CORALIE measurement, JD
2451185.1), there are five parameters, $P_j$, $K_j$, $e_j$,
$\omega_j$, and ${\cal M}_j$, for each Keplerian orbit.
Thus there are a total of 12 parameters.
The fact that the orbital parameters obtained from double-Keplerian
fits should be interpreted as Jacobi (and not astrocentric)
coordinates does not change the fitting function in equation
(\ref{RV}) but changes the conversion from $K_j$ and $P_j$ to the
planetary masses $m_j$ and orbital semimajor axes $a_j$:
\begin{equation}
K_j = \left(2 \pi G \over P_j\right)^{1/3} {m_j \sin i_j \over
      {m'_j}^{2/3}} {1 \over \sqrt{1 - e_j^2}} ,
\end{equation}
\begin{equation}
P_j = 2 \pi [a_j^3 / (G m'_j)]^{1/2} ,
\end{equation}
where ${m'_j} = \sum_{k=0}^j m_k$ \citep{lee03b}.
All of the orbital fits presented here are obtained using the
Levenberg-Marquardt algorithm \citep{pre92} to minimize $\chi_\nu^2$.
The Levenberg-Marquardt algorithm is effective at converging on a
local minimum for given initial values of the parameters.

\begin{figure}
\epsscale{0.7}
\plotone{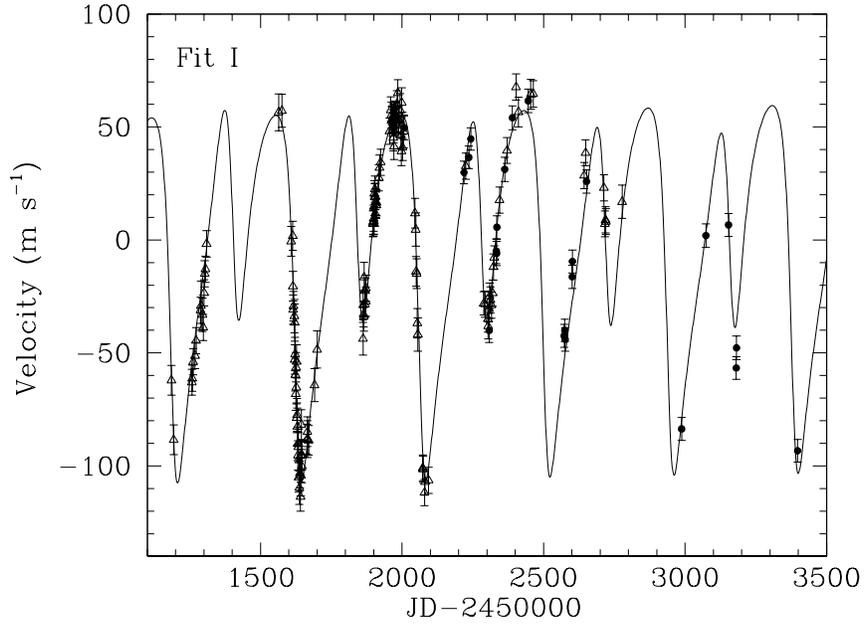}
\caption{
Radial velocity curve of the double-Keplerian fit I (with the minimum
$\chi_\nu^2$) compared to the Keck ({\it filled circles}) and CORALIE
({\it open triangles}) data for HD 82943.
The error bars show the quadrature sum of the internal uncertainties
and estimated stellar jitter ($4.2\mps$).
\label{fig:IRV}
}
\end{figure}

\begin{figure}
\epsscale{0.5}
\plotone{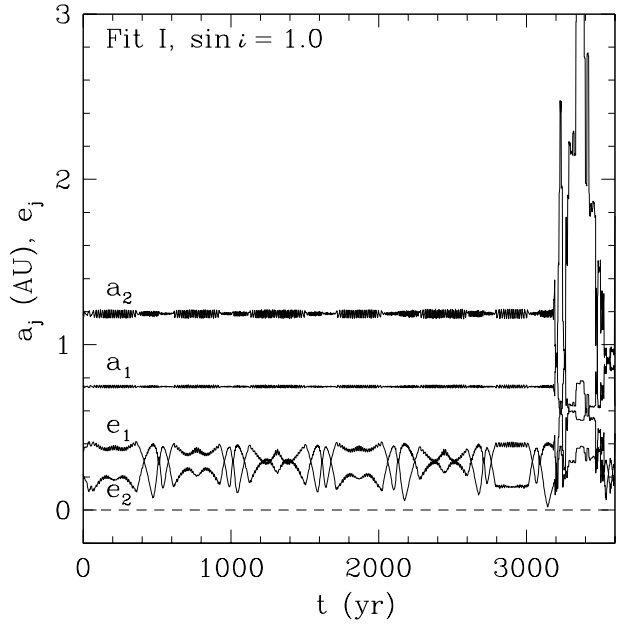}
\caption{
Evolution of the semimajor axes, $a_1$ and $a_2$, and eccentricities,
$e_1$ and $e_2$, for fit I with $\sin i = 1$ and starting epoch JD
2451185.1.
The system becomes unstable after $\sim 3200\yr$.
\label{fig:Ievol}
}
\end{figure}

We begin by using the parameter values listed in Table \ref{tab:mayor}
as the initial guess and allowing all 12 parameters to be varied in
the fitting.
The resulting fit I is listed in Table \ref{tab:fits} and compared to
the radial velocity data in Figure \ref{fig:IRV}.
We also show the residuals of the radial velocity data from fit I in
the top panel of Figure \ref{fig:res}.
Fit I has $\chi_\nu^2 = 1.87$ for 12 adjustable parameters (or $1.84$
if we rescale $\chi_\nu^2$ to 10 adjustable parameters for comparison
with the 10 parameter fits below) and RMS $= 7.88\mps$.
However, fit I is dynamically unstable if the orbits are assumed to be
coplanar.
We perform a series of direct numerical orbit integrations of fit I
similar to that described in \S 1 to test the stability of the
\citet{may04} best-fit solution (i.e., with $\sin i = 1$, $0.9$,
$\ldots$, $0.5$ and four starting epochs equally spaced between JD
2451185.1 and JD 2452777.7).
Figure \ref{fig:Ievol} shows the evolution of the semimajor axes,
$a_1$ and $a_2$, and eccentricities, $e_1$ and $e_2$, for the
integration with $\sin i = 1$ and starting epoch JD 2451185.1, which
becomes unstable after $\sim 3200\yr$.
All of the integrations of fit I become unstable in time ranging from
$\sim 60$ to $3200\yr$.

We then explore fits in which one or two of the parameters are fixed
at values different from those of fit I and the other parameters are
varied.
These experiments indicate that $e_2$ and $\omega_2$ are the least
constrained parameters, as there are fits with $e_2$ and $\omega_2$
fixed at values very different from those of fit I that have
$\chi_\nu^2$ (after taking into account the difference in the number
of adjustable parameters) and RMS only slightly different from those
of fit I.
Therefore, we decide to examine systematically best-fit
double-Keplerian models with fixed values of $e_2$ and $\omega_2$.
We start at $e_2 = 0.22$ and $\omega_2 = 290^\circ$ (which are close
to the values for fit I), using the parameters of fit I as the initial
guess for the other 10 parameters.
Then for each point on a grid of resolution $0.01$ in $e_2$ and $5^\circ$
in $\omega_2$, we search for the best fit with fixed $e_2$ and
$\omega_2$, using the best-fit parameters already determined for an
adjacent grid point as the initial guess.

Figure \ref{fig:param} shows the contours of $\chi_\nu^2$, RMS, $e_1$,
$\omega_1$, and the planetary mass ratio $m_1/m_2$ for the best-fit
double-Keplerian models with fixed values of $e_2$ and $\omega_2$.
Fit I at $e_2 = 0.219$ and $\omega_2 = 284^\circ$ is the only minimum
in $\chi_\nu^2$, with $\chi_\nu^2 = 1.84$ for 10 adjustable
parameters.
However, the minimum is relatively shallow, with $\chi_\nu^2 \la 2.0$
in a region that includes the entire region with $e_2 \la 0.2$ and
extends to $e_2 \approx 0.38$ near $\omega_2 = 130^\circ$ and
$280^\circ$.
The contours of RMS are similar to those of $\chi_\nu^2$.
The RMS minimum (where RMS $= 7.87\mps$) at $e_2 = 0.22$ and $\omega_2
= 275^\circ$ is only slightly offset from the $\chi_\nu^2$ minimum,
and the region with $\chi_\nu^2 \la 2.0$ also has RMS $\la 8.2\mps$.
The contours of $e_1$, $\omega_1$, and $m_1/m_2$ show that these
parameters are indeed well constrained, with $e_1$ within $\pm 0.05$
of $0.39$, $\omega_1$ within $\pm 15^\circ$ of $120^\circ$, and
$m_1/m_2$ within $\pm 0.2$ of $1.0$ for the fits with $\chi_\nu^2 \la
2.0$.

In Figure \ref{fig:param}{\it a} we mark the positions of three fits
II--IV (as well as fit I) whose dynamical properties are discussed in
detail in \S 4.
Their orbital parameters are listed in Table \ref{tab:fits}, and the
residuals of the radial velocity data from these fits are shown in the
lower three panels of Figure \ref{fig:res}.
We can infer from the parameters listed in Table \ref{tab:fits} that
the fits II--IV have nearly identical $a_1$($\approx 0.747\au$),
$a_2$($\approx 1.18\au$), and $m_2 \sin i$($\approx 1.71\,M_J$, where
$M_J$ is the mass of Jupiter) while $m_1 \sin i$ ranges from
$1.58\,M_J$ for fit II to $1.78\,M_J$ for fit IV.
Figure \ref{fig:res} demonstrates visually that these fits with
$\chi_\nu^2$ only slightly higher than the minimum are comparable in
the quality of fit to the fit I with the minimum $\chi_\nu^2$.

\section{DYNAMICAL ANALYSIS}

We begin with some general remarks on the dynamical properties of the
best-fit double-Keplerian models with fixed $e_2$ and $\omega_2$ that
can be inferred from the contour plots in Figure \ref{fig:param}.
There is a wide variety of stable 2:1 resonance configurations with
both $\theta_1$ and $\theta_2$ librating (see, e.g., \citealt{lee04}).
For $m_1/m_2 \sim 1$, there are (1) the sequence of resonance
configurations that can be reached by the differential migration of
planets with constant masses and initially coplanar and nearly
circular orbits, (2) the sequence with large $e_2$ and asymmetric
librations of both $\theta_1$ and $\theta_2$ far from either $0^\circ$
or $180^\circ$, and (3) the sequence with $\theta_1$ librating about
$180^\circ$ and $\theta_2$ librating about $0^\circ$.
Their loci in the $e_1$-$e_2$ plane are shown in Figures 5, 11, and
14, respectively, of \citet{lee04}.
The good fits in Figure \ref{fig:param} have $e_1 \approx 0.39 \pm
0.05$ and $e_2 \la 0.38$, which are far from the loci of the sequences
2 and 3.
Thus the good fits in Figure \ref{fig:param} cannot be
small-libration-amplitude configurations in the sequences 2 and 3.
Figure \ref{fig:param}{\it d} shows that the fits with $\omega_2
\approx 120^\circ$ also have $\omega_1 \approx 120^\circ$.
These fits with $\theta_{\rm SAR} = \varpi_1 - \varpi_2 \approx
0^\circ$ currently also have $e_1 \ga 0.40$ and $m_1/m_2 \la 1.0$, and
we can infer from Figure 5 of \citet{lee04} for the sequence 1 that
the fits with $e_2 \approx 0.14$ and $\omega_2 \approx 120^\circ$
could be in a symmetric configuration with both $\theta_1$ and
$\theta_2$ (and hence $\theta_{\rm SAR}$) librating about $0^\circ$
with small amplitudes.
However, direct numerical orbit integrations are needed to determine
whether the fits with $e_2 \approx 0.14$ and $\omega_2 \approx
120^\circ$ are in fact in such a resonance configuration.
Finally, Figure 5 of \citet{lee04} also shows that there are
configurations in the sequence 1 with asymmetric librations of
$\theta_1$ and $\theta_2$ in the region $e_1 \approx 0.39 \pm 0.05$
and $e_2 \la 0.38$ if $m_1/m_2 \ga 0.95$, and we cannot rule out the
possibility of asymmetric librations for the fits in Figure
\ref{fig:param} with $m_1/m_2 \ga 0.95$ without direct numerical orbit
integrations.

In order to use dynamical stability to narrow the range of reasonable
fits and to determine the actual dynamical properties of the stable
fits, we perform direct numerical orbit integrations of all of the
best-fit double-Keplerian models with fixed $e_2$ and $\omega_2$ shown
in Figure \ref{fig:param}.
We use the symplectic integrator SyMBA \citep{dun98} and assume that
the orbital parameters obtained from the fits are in Jacobi
coordinates (\citealt{lee03b}; see also \citealt{lis01}).
Double-Keplerian fits do not yield the inclinations of the orbits from
the plane of the sky.
We assume that the orbits are coplanar and consider two cases: $\sin i
= 1$ and $0.5$.
We perform integrations for only one starting epoch, JD 2451185.1,
since for each fit we already examine neighboring fits with slightly
different orbital parameters.
The maximum time interval of each integration is $5 \times 10^4\,$yr.
An integration is stopped and the model is considered to have become
unstable, if (1) the distance of a planet from the star is less than
$0.075\au$ or greater than $3.0\au$ or (2) there is an encounter
between the planets closer than the sum of their physical radii for an
assumed mean density of $1{\rm\,g\,cm}^{-3}$.
For these planets with orbital periods of about 220 and 440 days,
$a_1 \approx 0.75\au$, $a_2 \approx 1.2\au$, and their physical radii
are about 0.011 and 0.0066 of their Hill radii for a mean density of
$1{\rm\,g\,cm}^{-3}$.

In Figure \ref{fig:1.0} we present the results of the direct numerical
orbit integrations for $\sin i = 1$.
Figure \ref{fig:1.0}{\it a} shows the contours of the time,
$t_{\rm un}$, at which a model becomes unstable, and Figure
\ref{fig:1.0}{\it b} shows the contours of the semiamplitude
$\Delta\theta_1$.
All of the fits that are stable for at least $5 \times 10^4\yr$ have
$\theta_1$ librating about $0^\circ$ with $\Delta\theta_1 \la
140^\circ$.
Figure \ref{fig:1.0}{\it c} shows the contours of the semiamplitude
$\Delta\theta_2$, with the thick solid contour indicating the boundary
for the libration of $\theta_2$ about $0^\circ$.
None of the stable fits shows asymmetric libration of $\theta_1$ or
$\theta_2$.
Since the region with $\theta_2$ librating is smaller than that with
$\theta_1$ librating, there are stable good fits with only $\theta_1$
librating about $0^\circ$, in addition to stable good fits with both
$\theta_1$ and $\theta_2$ librating about $0^\circ$.
The fits with $e_2 \approx 0.14$ and $\omega_2 \approx 130^\circ$ have
small amplitude librations of both $\theta_1$ and $\theta_2$ about
$0^\circ$, in good agreement with the estimate of $e_2 \approx 0.14$
and $\omega_2 \approx 120^\circ$ above.

The contour for $\Delta\theta_2 = 30^\circ$ and the boundary for the
libration of $\theta_2$ are compared to the contour for $t_{\rm un} =
5 \times 10^4\yr$ and the contours for $\chi_\nu^2 = 1.85$ and $2.0$
in Figure \ref{fig:1.0}{\it d}, where the positions of the fits I--IV
are also indicated.
We can see from Figure \ref{fig:1.0}{\it d} that the fit I with the
minimum $\chi_\nu^2$, which is unstable (Fig. \ref{fig:Ievol}), is in
fact far from the stability boundary.
Fit II at $e_2 = 0.14$ and $\omega_2 = 130^\circ$ is the fit with the
smallest libration amplitudes of both $\theta_1$ and $\theta_2$.
The evolution of $a_j$, $e_j$, and $\theta_j$ for $2000\yr$ for fit II
with $\sin i = 1$ is shown in Figure \ref{fig:IIevol}.
The semiamplitude $\Delta\theta_1 \approx 6^\circ$ and
$\Delta\theta_2 \approx 10^\circ$, and the small amplitudes ensure
that both $a_j$ and $e_j$ have little variation.
Fit III at $e_2 = 0.02$ and $\omega_2 = 130^\circ$ is an example of a
stable fit with large amplitude librations of both $\theta_1$ and
$\theta_2$.
The semiamplitude $\Delta\theta_1 \approx 68^\circ$ and
$\Delta\theta_2 \approx 82^\circ$, the eccentricity $e_1$ varies
between $0.16$ and $0.40$ and $e_2$ between $0.02$ and $0.33$ (see
Fig. \ref{fig:IIIevol}).
Finally, fit IV at $e_2 = 0.02$ and $\omega_2 = 260^\circ$ is an
example of a stable fit with only $\theta_1$ librating.
Figure \ref{fig:IVevol} shows that $\theta_2$ circulates through $\pm
180^\circ$ when $e_2$ is very small.
The semiamplitude $\Delta\theta_1 \approx 94^\circ$, the eccentricity
$e_1$ varies between $0.08$ and $0.39$ and $e_2$ between $0.008$ and
$0.35$.

While the fits with small amplitude librations of both $\theta_1$ and
$\theta_2$ should be indefinitely stable, stability for $5 \times
10^4\yr$ does not in general guarantee stability on much longer
timescales.
Indeed the changes in the contours for $t_{\rm un} = 2000$, $10^4$,
and $5 \times 10^4\yr$ in Figure \ref{fig:1.0}{\it a} indicate that
the region of fits with only $\theta_1$ librating is likely to be
significantly smaller for stability on much longer timescales.
We confirm that the fit IV with only $\theta_1$ librating, as well as
fits II and III, are stable for at least $10^8\yr$ by performing
direct integrations of these fits for $10^8\yr$.
However, direct integrations of the fits at $e_2 = 0.02$ and $\omega_2
= 360^\circ$ and $310^\circ$ show that these fits become unstable
after $6.0 \times 10^6$ and $3.4 \times 10^7\yr$, respectively.

In Figure \ref{fig:0.5} we present the results of the direct numerical
orbit integrations for $\sin i = 0.5$.
The range of fits that are stable for at least $5 \times 10^4\yr$ is
significantly smaller than that for $\sin i = 1$ (compare
Figs. \ref{fig:1.0}{\it a} and \ref{fig:0.5}{\it a}).
There are small regions where the fits have only $\theta_1$
librating about $0^\circ$ and they are stable for $5 \times 10^4\yr$
(see Fig. \ref{fig:0.5}{\it d}), but the changes in the contours for
$t_{\rm un} = 2000$, $10^4$, and $5 \times 10^4\yr$ in Figure
\ref{fig:0.5}{\it a} indicate that these fits are unlikely to be
stable on much longer timescales.
Thus, for $\sin i = 0.5$, only fits with both $\theta_1$ and
$\theta_2$ librating about $0^\circ$ are stable.
The unstable fit I is again far from the stability boundary.
Fit IV, which is stable for at least $10^8\yr$ for $\sin i = 1$, is
unstable for $\sin i = 0.5$.
Direct integrations for $10^8\yr$ show that fits II and III with $\sin
i = 0.5$ are stable for at least $10^8\yr$.
We can see from Figures \ref{fig:1.0}{\it b} and \ref{fig:0.5}{\it b}
and Figures \ref{fig:1.0}{\it c} and \ref{fig:0.5}{\it c} that there
are only small changes in the contours of the semiamplitudes
$\Delta\theta_1$ and $\Delta\theta_2$ for the fits that are stable for
both $\sin i = 0.5$ and $1$.
This means that the changes in $\Delta\theta_1$ and $\Delta\theta_2$
with $\sin i$ are small for most of the fits that are stable for both
$\sin i = 0.5$ and $1$.
In particular, fit II is also the fit with the smallest libration
amplitudes of both $\theta_1$ and $\theta_2$ for $\sin i = 0.5$.
Figure \ref{fig:IIevol2} (compared to Fig. \ref{fig:IIevol}) shows
that the libration period is shorter due to the larger planetary
masses but that $\Delta\theta_1$ and $\Delta\theta_2$ are within
$1^\circ$ of the values for $\sin i = 1$.
We do not show the time evolution of $\theta_1$ and $\theta_2$ for fit
III with $\sin i = 0.5$, but $\Delta\theta_1$ and $\Delta\theta_2$ are
again within $1^\circ$ of the values for $\sin i = 1$.

\section{DISCUSSION AND CONCLUSIONS}

We have analyzed the HD 82943 planetary system by examining the
best-fit double-Keplerian model to the radial velocity data as a
function of the poorly constrained eccentricity and argument of
periapse of the outer planet's orbit.
We have not found any good fits that are non-resonant and dynamically
stable (if the orbits are assumed to be coplanar), and the two planets
in the HD 82943 system are almost certainly in 2:1 mean-motion
resonance.
The fit I with the minimum $\chi_{\nu}^2$ is unstable, but there is a
wide range of fits outside the minimum with $\chi_{\nu}^2$ only
slightly higher than the minimum.
If the unknown $\sin i \approx 1$, there are stable good fits with
both of the mean-motion resonance variables, $\theta_1 = \lambda_1 -
2\lambda_2 + \varpi_1$ and $\theta_2 = \lambda_1 - 2\lambda_2 +
\varpi_2$, librating about $0^\circ$ (e.g., fits II and III), as well
as stable good fits with only $\theta_1$ librating about $0^\circ$
(e.g., fit IV).
If $\sin i \approx 0.5$, only good fits with both $\theta_1$ and
$\theta_2$ librating about $0^{\circ}$ (e.g., fits II and III) are
stable.
Fit II is the fit with the smallest libration semiamplitudes of both
$\theta_1$ and $\theta_2$, with $\Delta\theta_1 \approx 6^\circ$ and
$\Delta\theta_2 \approx 10^\circ$.

Our analysis differs from that of \citet{fmb05} in having the
additional Keck data, in using $\chi_\nu^2$ instead of RMS as the
primary measure of the goodness of fit, and in examining
systematically the dynamical properties of all of the fits found.
Although \citet{fmb05} showed $e_j$ and $\omega_j$ of many fits with
RMS within $\approx 0.2\mps$ of the minimum RMS, they discussed in
detail the dynamical properties of only two fits --- their solutions A
and B --- with RMS within $0.06\mps$ of the minimum.
Their solutions A and B have $\omega_1 \approx 120^\circ$ and
$\omega_2 \ga 200^\circ$ and hence large amplitude librations of both
$\theta_1$ and $\theta_2$.
Because the fit to the CORALIE data with the minimum RMS (which is
near the minimum $\chi_\nu^2$ fit of \citealt{may04} at $e_2 = 0.18$
and $\omega_2 = 237^\circ$; see Table \ref{tab:mayor}) is close to the
stability boundary in the $e_2$-$\omega_2$ plane, \citet{fmb05} were
able to find stable fits with RMS within $0.06\mps$ of the minimum.
For the combined data set, the fit with the minimum RMS at $e_2 =
0.22$ and $\omega_2 = 275^\circ$ is far from the stability boundary,
and none of our fits with RMS within $0.06\mps$ of the minimum are
stable (see Figs. \ref{fig:param}{\it b}, \ref{fig:1.0}{\it a}, and
\ref{fig:0.5}{\it a}).
However, by examining systematically the $\chi_\nu^2$, RMS, and
dynamical properties of all of our fits, we have found fits like fit
IV (which has stable, large amplitude libration of only $\theta_1$ if
$\sin i \approx 1.0$), fit III (which has large amplitude librations
of both $\theta_1$ and $\theta_2$) and, in particular, fit II (which
has small amplitude librations of both $\theta_1$ and $\theta_2$), all
with RMS within $0.13\mps$ of the minimum and $\chi_\nu^2$ within
$0.08$ of the minimum.
The stable and unstable regions in the $e_2 \cos \omega_2$-$e_2 \sin
\omega_2$ plane shown in Figures 4 and 8 of \citet{fmb05} are
qualitatively similar to those in our Figures \ref{fig:1.0}{\it a} and
\ref{fig:0.5}{\it a} in the $e_2$-$\omega_2$ plane, but it should be
noted that their figures show simulations of their solution B (for the
CORALIE data) with $e_2$ and $\omega_2$ changed, while our figures
show simulations of the fits (to the combined data set) that minimize
$\chi_\nu^2$ for the given $e_2$ and $\omega_2$.

The relatively large values of $\chi_\nu^2$ ($\ge 1.84$) and RMS ($\ge
7.87\mps$) of the fits presented in this paper suggest that the
double-Keplerian model may not fully explain the radial velocity data
of HD 82943.
Also hinting at the same possibility is the increase in the RMS of the
best fit from $6.99\mps$ (for the extracted CORALIE data alone) to
$7.88\mps$ with the inclusion of the Keck data, which fill in some
gaps in the CORALIE data and increase the time span of observations.
However, the RMS and $\chi_\nu^2$ values are by themselves not very
strong evidence, because our estimate ($4.2\mps$) for the radial
velocity jitter has an uncertainty of $\sim 50\%$ and the jitter can
be $\sim 6\mps$.
On the other hand, there appears to be systematic deviations of the
radial velocity data from the double-Keplerian fits.
Figure \ref{fig:res} shows that the data within about $\pm 200$ days
of JD 2452600 are higher than the fits and that the data near JD
2453181 are lower than the fits.
The fact that $\sim 40\%$ of the Keck measurements fall in these two
regions result in the aforementioned increase in the RMS of the best
fit from $6.99\mps$ for the CORALIE data alone to $7.88\mps$ for the
combined data set.
But it is important to note that the Keck and CORALIE data are
consistent with each other where they overlap and that they are both
higher than the fits in the region around JD 2452600.
One possible explanation for the deviations is that they are due to
the mutual gravitational interaction of the planets.
However, unlike the GJ 876 system where the resonance-induced apsidal
precession rate $d\varpi_j/dt$ is $-41^\circ \yr^{-1}$ on average and
the precession has been seen for more than one full period
\citep{lau05}, the average apsidal precession rates of, e.g., fits II
and III are only $-0\fdg51\,(\sin i)^{-1} \yr^{-1}$ and
$-0\fdg71\,(\sin i)^{-1} \yr^{-1}$, respectively, and the orbits have
precessed a negligible $3$--$4\,(\sin i)^{-1}$ degrees over the
$6.1\yr$ time span of the available observations.
On the other hand, there are small variations in the orbital elements
on shorter timescales due to the planetary interaction that could
produce deviations from the double-Keplerian model.
Alternatively, the deviations could be due to the presence of
additional planet(s).
Continued observations of HD 82943 combined with dynamical fits should
allow us to distinguish these possibilities.

\acknowledgments
It is a pleasure to thank Stan Peale for many informative discussions.
We also thank C. Beaug\'e and the referee for their comments on the
manuscript.
This research was supported in part by NASA grants NAG5-11666 and
NAG5-13149 (M.H.L.), by NASA grant NAG5-12182 and travel support from
the Carnegie Institution of Washington (R.P.B.), by NASA grant
NAG5-75005 (G.W.M.), and by NSF grant AST-0307493 (S.S.V.).

\clearpage

\begin{deluxetable}{lcccc}
\tablecolumns{5}
\tablewidth{0pt}
\tablecaption{Orbital Parameters of the HD 82943 Planets from \citet{may04}
\label{tab:mayor}}
\tablehead{
\colhead{Parameter} & \colhead{} & \colhead{Inner} & \colhead{} &
\colhead{Outer}
}
\startdata
$P$ (days) & & 219.4 & & 435.1 \\
$K$ (${\rm m}\,{\rm s}^{-1}$) & & 61.5 & & 45.8 \\
$e$ & & 0.38 & & 0.18 \\
$\omega$ (deg) & & 124 & & 237 \\
${\cal M}$ (deg) & & 357 & & 246 \\
\enddata
\tablecomments{The parameters are the orbital period $P$, the velocity
amplitude $K$, the orbital eccentricity $e$, the argument of
periapse $\omega$, and the mean anomaly ${\cal M}$ at the epoch JD
2451185.1.}
\end{deluxetable}

\clearpage

\begin{deluxetable}{rrrrr}
\tablecolumns{5}
\tablewidth{0pt}
\tablecaption{Measured Velocities for HD 82943 from Keck
\label{tab:RV}}
\tablehead{
\colhead{JD} & \colhead{} & \colhead{Radial Velocity} & \colhead{} &
\colhead{Uncertainty} \\
\colhead{($-$2450000)} & \colhead{} &
\colhead{(${\rm m}\,{\rm s}^{-1}$)} & \colhead{} &
\colhead{(${\rm m}\,{\rm s}^{-1}$)}
}
\startdata
2006.913 & &     43.39 & &     3.2 \\
2219.121 & &     23.58 & &     2.7 \\
2236.126 & &     30.29 & &     2.7 \\
2243.130 & &     38.46 & &     2.6 \\
2307.839 & &  $-$46.33 & &     3.2 \\
2332.983 & &  $-$11.26 & &     3.4 \\
2333.956 & &  $-$12.25 & &     3.3 \\
2334.873 & &   $-$0.65 & &     3.0 \\
2362.972 & &     25.02 & &     3.3 \\
2389.944 & &     47.81 & &     3.1 \\
2445.739 & &     55.26 & &     3.0 \\
2573.147 & &  $-$48.70 & &     2.9 \\
2575.140 & &  $-$46.26 & &     2.6 \\
2576.144 & &  $-$50.28 & &     3.1 \\
2601.066 & &  $-$22.60 & &     3.1 \\
2602.073 & &  $-$15.76 & &     2.9 \\
2652.001 & &     19.60 & &     3.1 \\
2988.109 & &  $-$89.93 & &     2.8 \\
3073.929 & &   $-$4.31 & &     3.1 \\
3153.754 & &      0.41 & &     2.8 \\
3180.745 & &  $-$62.93 & &     2.7 \\
3181.742 & &  $-$54.04 & &     3.1 \\
3397.908 & &  $-$99.57 & &     2.8 \\
\enddata
\end{deluxetable}

\clearpage

\begin{deluxetable}{lcccc}
\tablecolumns{5}
\tablewidth{0pt}
\tablecaption{Orbital Parameters of the HD 82943 Planets
\label{tab:fits}}
\tablehead{
\colhead{Parameter} & \colhead{} & \colhead{Inner} & \colhead{} &
\colhead{Outer}
}
\startdata
\multicolumn{5}{c}{Fit I [$\chi_\nu^2 = 1.87$, $\chi_\nu^2(10) = 1.84$, RMS $= 7.88\mps$]} \\*
\noalign{\vskip .7ex}
\cline{1-5}
\noalign{\vskip .7ex}
$P$ (days) & & 219.3 & & 441.2 \\*
$K$ (${\rm m}\,{\rm s}^{-1}$) & & 66.0 & & 43.6 \\*
$e$ & & 0.359 & & 0.219 \\*
$\omega$ (deg) & & 127 & & 284 \\*
${\cal M}$ (deg) & & 353 & & 207 \\*
$m \sin i$ ($M_J$) & & 2.01 & & 1.75 \\*
$V_K$ (${\rm m}\,{\rm s}^{-1}$) & \multicolumn{4}{c}{-6.3} \\*
$V_C$ (${\rm m}\,{\rm s}^{-1}$) & \multicolumn{4}{c}{8143.9} \\
\noalign{\vskip .7ex}
\cline{1-5}
\noalign{\vskip .7ex}
\multicolumn{5}{c}{Fit II ($\chi_\nu^2 = 1.92$, RMS $= 8.00\mps$)} \\*
\noalign{\vskip .7ex}
\cline{1-5}
\noalign{\vskip .7ex}
$P$ (days) & & 219.6 & & 436.7 \\*
$K$ (${\rm m}\,{\rm s}^{-1}$) & & 53.5 & & 42.0 \\*
$e$ & & 0.422 & & 0.14 \\*
$\omega$ (deg) & & 122 & & 130 \\*
${\cal M}$ (deg) & & 357 & & 352 \\*
$m \sin i$ ($M_J$) & & 1.58 & & 1.71 \\*
$V_K$ (${\rm m}\,{\rm s}^{-1}$) & \multicolumn{4}{c}{-6.9} \\*
$V_C$ (${\rm m}\,{\rm s}^{-1}$) & \multicolumn{4}{c}{8142.7} \\
\noalign{\vskip .7ex}
\cline{1-5}
\noalign{\vskip .7ex}
\multicolumn{5}{c}{Fit III ($\chi_\nu^2 = 1.91$, RMS $= 7.99\mps$)} \\*
\noalign{\vskip .7ex}
\cline{1-5}
\noalign{\vskip .7ex}
$P$ (days) & & 219.5 & & 438.8 \\*
$K$ (${\rm m}\,{\rm s}^{-1}$) & & 58.1 & & 41.7 \\*
$e$ & & 0.398 & & 0.02 \\*
$\omega$ (deg) & & 120 & & 130 \\*
${\cal M}$ (deg) & & 357 & & 356 \\*
$m \sin i$ ($M_J$) & & 1.74 & & 1.71 \\*
$V_K$ (${\rm m}\,{\rm s}^{-1}$) & \multicolumn{4}{c}{-6.3} \\*
$V_C$ (${\rm m}\,{\rm s}^{-1}$) & \multicolumn{4}{c}{8143.2} \\
\noalign{\vskip .7ex}
\cline{1-5}
\multicolumn{5}{c}{} \\
\multicolumn{5}{c}{} \\
\multicolumn{5}{c}{} \\
\multicolumn{5}{c}{} \\
\multicolumn{5}{c}{} \\
\multicolumn{5}{c}{} \\
\multicolumn{5}{c}{\phantom{Fit I [$\chi_\nu^2 = 1.87$, $\chi_\nu^2(10) = 1.84$, RMS $= 7.88\mps$]}} \\
\noalign{\vskip -3ex}
\multicolumn{5}{c}{Fit IV ($\chi_\nu^2 = 1.90$, RMS $= 7.97\mps$)} \\*
\noalign{\vskip .7ex}
\cline{1-5}
\noalign{\vskip .7ex}
$P$ (days) & & 219.5 & & 439.2 \\*
$K$ (${\rm m}\,{\rm s}^{-1}$) & & 59.3 & & 41.7 \\*
$e$ & & 0.391 & & 0.02 \\*
$\omega$ (deg) & & 121 & & 260 \\*
${\cal M}$ (deg) & & 356 & & 227 \\*
$m \sin i$ ($M_J$) & & 1.78 & & 1.72 \\*
$V_K$ (${\rm m}\,{\rm s}^{-1}$) & \multicolumn{4}{c}{-6.2} \\*
$V_C$ (${\rm m}\,{\rm s}^{-1}$) & \multicolumn{4}{c}{8143.3} \\
\enddata
\tablecomments{Fits II--IV are 10 parameter fits with fixed
$e$ and $\omega$ of the outer planet's orbit.
Fit I is a 12 parameter fit, and its $\chi_\nu^2$ is shown
both at its true value and at the value rescaled to 10 adjustable
parameters for comparison with the 10 parameter fits.
In addition to the parameters $P$, $K$, $e$, $\omega$, and ${\cal M}$,
the minimum planetary mass, $m \sin i$, and the zero-point velocities,
$V_K$ and $V_C$, of the Keck and CORALIE data, respectively, are
listed.}
\end{deluxetable}

\clearpage

\clearpage

\begin{figure}
\epsscale{0.9}
\plotone{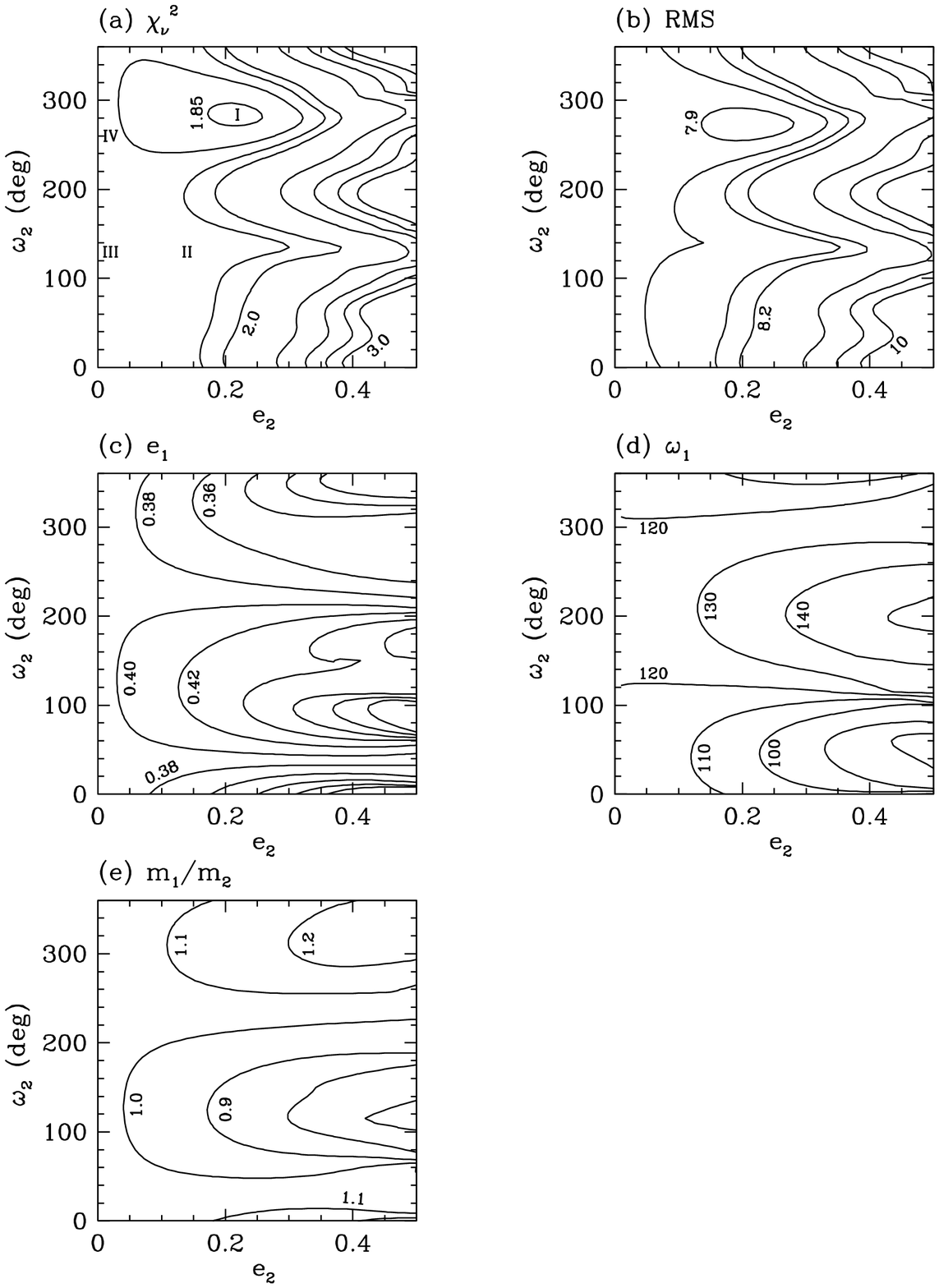}
\end{figure}

\clearpage

\begin{figure}
\caption{
Parameters of the best-fit double-Keplerian models for HD 82943 for
fixed values of the eccentricity, $e_2$, and argument of periapse,
$\omega_2$, of the outer planet's orbit.
({\it a}) Contours of $\chi_\nu^2$ in steps of 0.05 from 1.85 to 2.0
and then in steps of 0.25 to 3.0.
The positions of fits I--IV are marked.
({\it b}) Contours of the RMS of the residuals (in units of ${\rm
  m}\,{\rm s}^{-1}$) in steps of 0.1 from 7.9 to 8.2 and then in steps
of 0.6 to 10.
({\it c}) Contours of the eccentricity, $e_1$, of the inner orbit in
steps of 0.02 from 0.3 to 0.5.
({\it d}) Contours of the argument of periapse, $\omega_1$, of the
inner orbit in steps of $10^\circ$ from $80^\circ$ to $150^\circ$.
({\it e}) Contours of the planetary mass ratio $m_1/m_2$ in steps of
0.1 from 0.7 to 1.2.
The ratio $m_1/m_2$ is nearly independent of $\sin i $, as long as
$\sin i$ is not very small and $m_1, m_2 \ll m_0$.
\label{fig:param}
}
\end{figure}

\begin{figure}
\epsscale{0.7}
\plotone{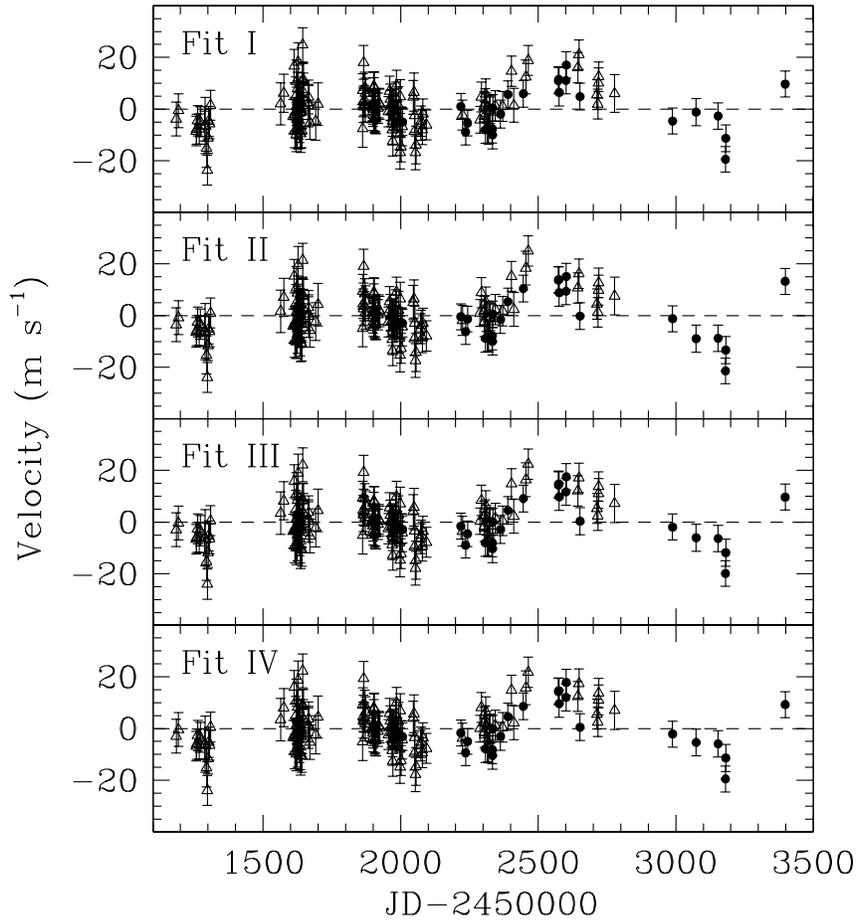}
\caption{
Residual velocities from fits I--IV of the Keck ({\it filled
circles}) and CORALIE ({\it open triangles}) data for HD 82943.
The error bars show the quadrature sum of the internal uncertainties
and estimated stellar jitter ($4.2\mps$).
\label{fig:res}
}
\end{figure}

\clearpage

\begin{figure}
\epsscale{0.9}
\plotone{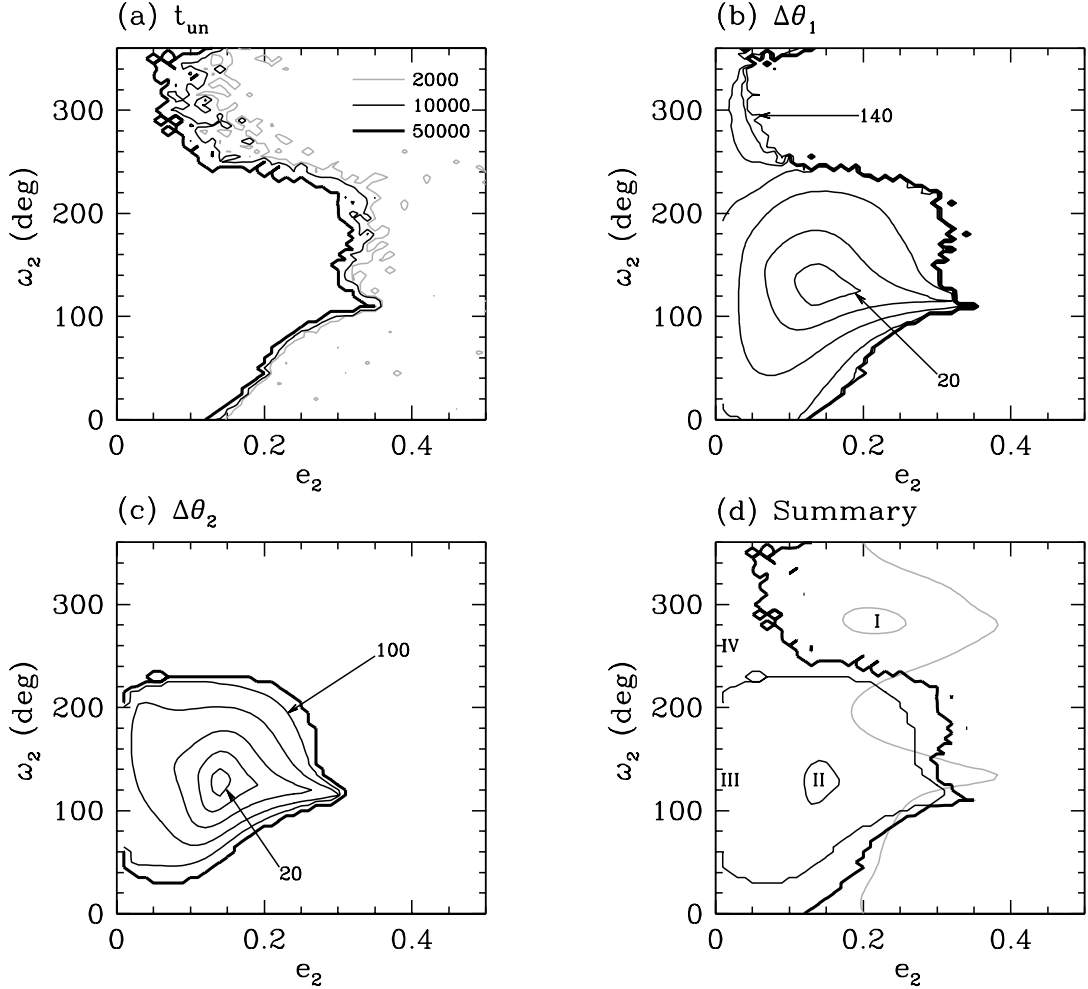}
\caption{
Dynamical properties of the best-fit double-Keplerian models shown in
Fig. \ref{fig:param} for $\sin i = 1$.
({\it a}) Contours of the time $t_{\rm un}$ (in units of yr) at which
a model becomes unstable.
({\it b}) Contours of the semiamplitude $\Delta\theta_1$ in steps of
$20^\circ$.
({\it c}) Contours of the semiamplitude $\Delta\theta_2$ in steps of
$20^\circ$ from $20^\circ$ to $100^\circ$.
The thick solid contour is the boundary for the libration of
$\theta_2$.
({\it d}) Positions of fits I--IV in the $e_2$-$\omega_2$ plane.
The gray contours are $\chi_\nu^2 = 1.85$ and $2.0$.
The thick solid contour is $t_{\rm un} = 5 \times 10^4\yr$.
The thin solid contours are $\Delta\theta_2 = 30^\circ$ and the
boundary for the libration of $\theta_2$.
Only fits with $\Delta\theta_1 \la 140^\circ$ are stable for at least
$5 \times 10^4\yr$.
Fit I at the minimum of $\chi_\nu^2$ is unstable.
Fit II is the stable fit with the smallest libration amplitudes of both
$\theta_1$ and $\theta_2$.
Fit III is a stable fit with large amplitude librations of both
$\theta_1$ and $\theta_2$.
Fit IV is a stable fit with only $\theta_1$ librating.
\label{fig:1.0}
}
\end{figure}

\clearpage

\begin{figure}
\epsscale{0.48}
\plotone{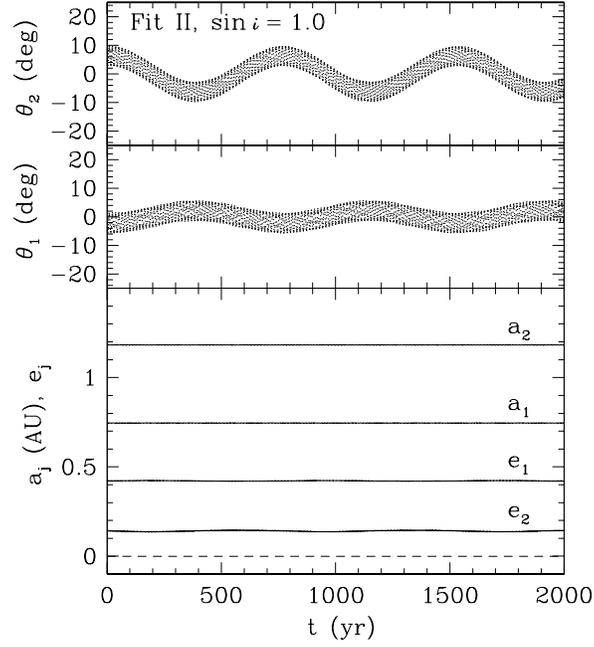}
\caption{
Evolution of the semimajor axes, $a_1$ and $a_2$, eccentricities,
$e_1$ and $e_2$, and 2:1 mean-motion resonance variables, $\theta_1 =
\lambda_1 - 2 \lambda_2 + \varpi_1$ and $\theta_2 = \lambda_1 - 2
\lambda_2 + \varpi_2$, for fit II with $\sin i = 1$.
The semiamplitude $\Delta\theta_1 \approx 6^\circ$ and $\Delta\theta_2
\approx 10^\circ$.
\label{fig:IIevol}
}
\end{figure}

\begin{figure}
\epsscale{0.48}
\plotone{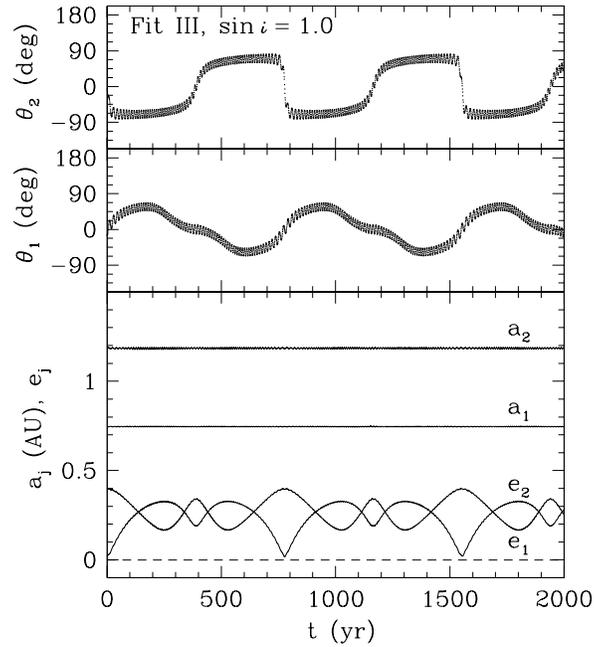}
\caption{
Same as Fig. \ref{fig:IIevol}, but for fit III with $\sin i = 1$.
The semiamplitude $\Delta\theta_1 \approx 68^\circ$ and
$\Delta\theta_2 \approx 82^\circ$.
\label{fig:IIIevol}
}
\end{figure}

\clearpage

\begin{figure}
\epsscale{0.48}
\plotone{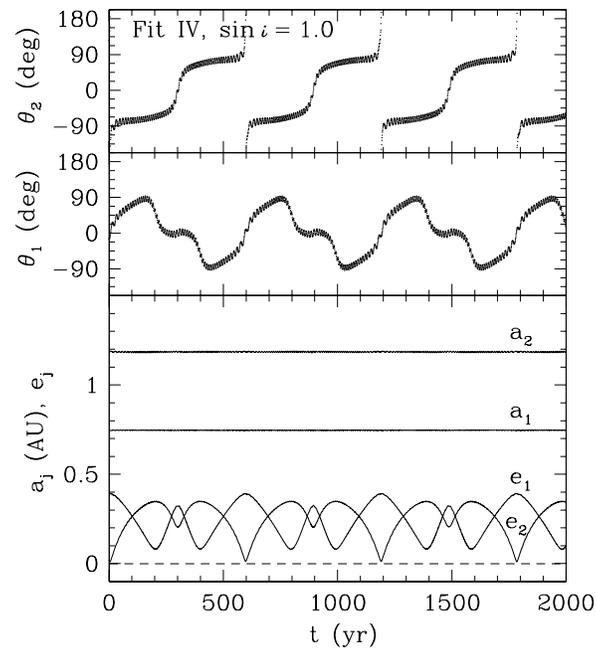}
\caption{
Same as Fig. \ref{fig:IIevol}, but for fit IV with $\sin i = 1$.
The semiamplitude $\Delta\theta_1 \approx 94^\circ$, and $\theta_2$
circulates.
\label{fig:IVevol}
}
\end{figure}

\clearpage

\begin{figure}
\epsscale{0.9}
\plotone{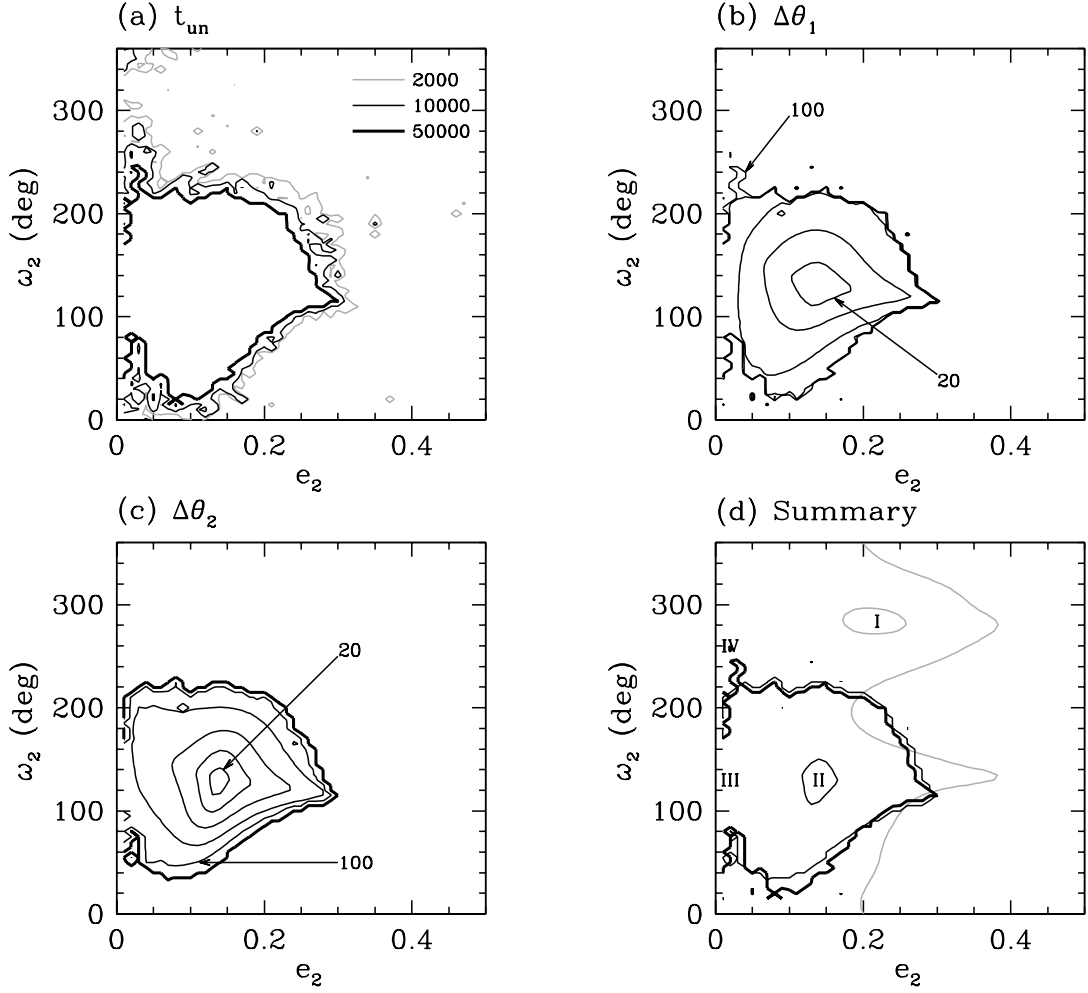}
\caption{
Same as Fig. \ref{fig:1.0}, but for $\sin i = 0.5$.
Only fits with both $\theta_1$ and $\theta_2$ librating are stable,
and fits I and IV are unstable.
\label{fig:0.5}
}
\end{figure}

\clearpage

\begin{figure}
\epsscale{0.48}
\plotone{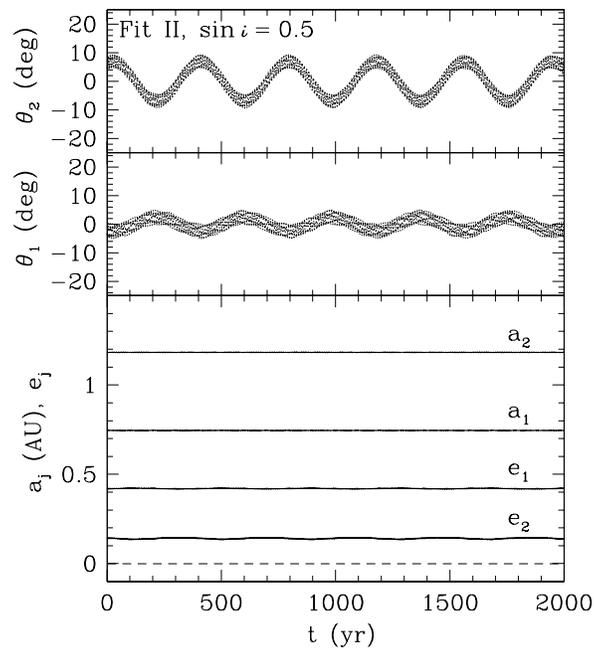}
\caption{
Same as Fig. \ref{fig:IIevol}, but for fit II with $\sin i = 0.5$.
The semiamplitudes $\Delta\theta_1$ and $\Delta\theta_2$ are within
$1^\circ$ of the values for $\sin i = 1$.
\label{fig:IIevol2}
}
\end{figure}

\end{document}